# Evidence for the Meissner effect in the nickelate superconductor $La_3Ni_2O_{7-\delta}$ single crystal using diamond quantum sensors

Lin Liu[1,3,†], Jianning Guo[2,†], Deyuan Hu[4], Guizhen Yan[1,3], Yuzhi Chen[2], Lunxuan Yu[1,3], Meng Wang[4,*],

Xiao-Di Liu[1,*] and Xiaoli Huang[2,*]

[1]*Key Laboratory of Materials Physics, Institute of Solid State Physics, HFIPS, Chinese Academy of Sciences, Hefei 230031, China*
[2] *State Key Laboratory of High Pressure and Superhard Materials, College of Physics, Jilin University, Changchun 130012, China*
[3]*University of Science and Technology of China, Hefei 230026, China*
[4]*Guangdong Provincial Key Laboratory of Magnetoelectric Physics and Devices, School of Physics, Sun Yat-Sen University, Guangzhou 510275, China.*

[†]These authors contributed equally

[*]**Corresponding emails:** wangmeng5@mail.sysu.edu.cn; xiaodi@issp.ac.cn; huangxiaoli@jlu.edu.cn

**Abstract:**

Quantum sensing with nitrogen-vacancy (NV) centers in diamond enables the characterization of magnetic properties in the extreme situation of tiny sample with defects. Recent studies have reported superconductivity in $La_3Ni_2O_{7-\delta}$ under pressure, with zero-resistance near 80 K, though the Meissner effect remains debated due to low superconducting volume fractions and limited high-pressure magnetic measurement techniques. In this work, we use diamond quantum sensors and four-probe detection to observe both zero resistance and the Meissner effect in the same $La_3Ni_2O_{7-\delta}$ single crystal. By mapping the Meissner effect, we visualized superconducting regions and revealed sample inhomogeneities. Our combined magnetic and electrical measurements on the same crystal provide dual evidence of superconductivity, supporting the high-temperature superconductivity of $La_3Ni_2O_{7-\delta}$. This study also offers insights into its structural and magnetic properties under high pressure.

Unconventional superconductors have attracted great attentions since the discovery of high-temperature superconductivity in cuprates. As the first family of unconventional superconducting materials with superconducting transition temperature ($T_c$) above the boiling point of liquid nitrogen, cuprates are characterized by $CuO_2$ layers, a rich phase diagram and superconducting fluctuations [1-4]. In the cuprates system, superconductivity occurs within the $CuO_2$ sheets with weak interlayer coupling [5-8]. Despite extensive research, the possible pairing mechanisms in cuprates remain debate, driving the search for new families of unconventional superconductors to uncover fundamental insights into this phenomenon.

Nickelates have emerged as a promising platform for studying unconventional superconductivity due to their proximity to copper on the periodic table, suggesting potential similarities. However, superconductivity in nickelates was only observed in 2019, with a relatively $T_c$ of 5–20 K in infinite-layer nickelate thin films [9-12]. Recently, Sun et al. reported a significant breakthrough with a $T_c$ close to 80 K in the Ruddlesden–Popper bilayer perovskite $La_3Ni_2O_7$ above 14 GPa [13]. Unlike infinite-layer nickelates and cuprates, $La_3Ni_2O_7$ features a unique bilayer $NiO_2$ lattice with $Ni^{2.5+}$ ions [14], distinguishing it structurally and electronically. Using a liquid pressure-transmitting medium enabled zero resistance—a hallmark of superconductivity [15,16]. However, there are debates about the nature of the superconductivity in compressed $La_3Ni_2O_7$. Some suggests it may be filamentary, with a low superconducting volume fraction [17], raising the critical question of whether $La_3Ni_2O_7$ supports bulk superconductivity, which requires confirmation through the Meissner effect—evidence of magnetic field expulsion. Demonstrating both zero resistance and the Meissner effect in the same sample is crucial for confirming superconductivity, especially in micrometer-scale samples, posing a fundamental challenge for the reliable studies.

To induce high-temperature superconductivity, the $La_3Ni_2O_{7-\delta}$ single-crystal samples are imposed in diamond anvil cells, in which the sample dimension is restricted in hundreds of microns. Additionally, sample and pressure heterogeneities often result in superconductivity occurring in only small regions of the sample. In these harsh experimental conditions, credible techniques are required to detect the Meissner effect.

Quantum sensing with nitrogen-vacancy (NV) centers in diamond enables the characterization of magnetic properties even in the extreme situation of tiny samples with defects [18-22], particularly suitable for simultaneous measurement of zero resistance and the Meissner effect in the same sample. This method has been proven effective in different types of superconductors, including cuprates [20], $MgB_2$ [18,19], and hydride $CeH_9$ [23].

The NV center in a diamond is a point defect with a ground-state electron spin Hamiltonian [24] expressed as $H = DS_z^2 + E(S_x^2 - S_y^2) - \gamma_e \boldsymbol{B} \cdot \boldsymbol{S}$, where D and E are zero-field splitting parameters determined by the diamond lattice field. At room temperature and atmospheric pressure, $D = 2.87$ GHz, and $E = 0$ in the absence of electric field and stress. Here $\boldsymbol{B}$ represents the applied magnetic field, and $\boldsymbol{S} = (S_x, S_y, S_z)$ denotes the NV electron spin [see Figure 1(c)]. The electron gyromagnetic ratio is $\gamma_e = -2.8$ MHz/G. The NV center's properties can be modulated by magnetic field [25,26], pressure [27,28], temperature [29] and electric field [30]. In this study, since no electric field is applied, its effect on the NV center is disregarded. Additionally, temperature effects on the central resonance frequency are negligible compared to the contributions from magnetic field and pressure. Considering that the nickel-based superconducting transition temperature are below 100 K, and the D-value of the NV center remains stable in this range [31], temperature influence is also negligible. Thus, the splitting observed in ODMR spectra is primarily attributed to magnetic field and stress.

In this study, we combined diamond quantum sensors with a four-probe detection method to simultaneously observe both zero resistance and the Meissner effect in the same single-crystal sample of $La_3Ni_2O_{7-\delta}$. By locally mapping the Meissner effect across different sample regions, we revealed dual superconducting evidence in the same sample. NV-center quantum sensing with high resolution can accurately detect the spatial diamagnetism of the samples. This approach provides critical evidence supporting high-temperature superconductivity in $La_3Ni_2O_{7-\delta}$ and deepens our understanding of unconventional superconductivity in nickelates.

As shown in Figure 1(a), the sample is positioned between two diamond anvils to generate high-pressure condition, enabling in-situ electrical transport and ODMR measurements. To obtain better hydrostatic pressure and zero resistance, ammonia borane is applied as the pressure-transmitting medium. To probe the variations of the local magnetic field as the sample transitions into the superconducting state, carbon ion implantation and annealing are conducted on the (111)-oriented type Ib diamond anvils to create NV centers [see Figure 1(b)], as reported in previous work [22,23]. The detailed experimental procedures are provided in the Methods section.

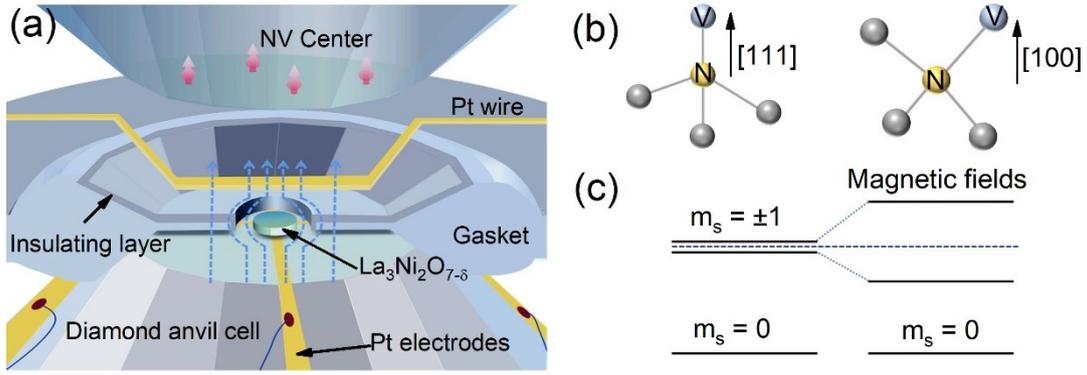

FIG. 1. The NV centers in diamond anvil cell (DAC) under high pressure. (a) The schematic of DAC with NV centers embedded in the diamond culet. The $La_3Ni_2O_{7-\delta}$ sample is loaded into the chamber. For transferring the microwave signal, a Pt wire is placed across the diamond culet. (b) NV centers aligned along the positive axis of the coordinate system. The NV centers in diamonds which polished along different crystal orientations, left: [111], right: [100]. (c) The spin energy structure of NV centers under applied magnetic fields. The axial magnetic fields induce the Zeeman splitting of the $m_s = \pm 1$ states.

Figure 2(a) shows the results of the four-probe electrical transport measurements. The $La_3Ni_2O_{7-\delta}$ sample exhibits superconducting transitions at 72 K at 22 GPa and 65 K at 28 GPa, respectively, which is consistent with the previous work [13,15,16]. The zero resistance is achieved below 8.5 K at 22 GPa. The applied magnetic fields of 0–8 T further prove the superconducting transition as shown in Figures 2(b) and 2(c). We determine the upper critical magnetic field ($\mu_0 H_{c2}$) of 86 T at 22 GPa and 71 T at 28 GPa, by using the Ginzburg-Landau (GL) [32] equations, respectively [see Figure 2(d)].

After confirming the superconducting zero resistance of the single-crystal sample,

we conducted magnetic measurement using an NV center quantum sensor in diamond. The NV quantum sensor was integrated into a cryogenic system [33,34] equipped with a scanning confocal microscope, enabling the investigation of local magnetic properties. ODMR spectroscopy, based on NV centers, was performed at various points within the sample chamber to detect the local magnetic response of $La_3Ni_2O_{7-\delta}$. Under an external magnetic field, Zeeman splitting occurs in the NV center's spin level [see Figure 1(c)], resulting in the splitting of ODMR spectra. The degree of splitting is directly proportional to the strength of the magnetic field, allowing calculation of the local magnetic field magnitude from the splitting values. This technique, which has been extensively in DAC [18-20,23] under pressure reaching the megapascal range [23,35], was used to detect local diamagnetism by cooling the sample below $T_c$.

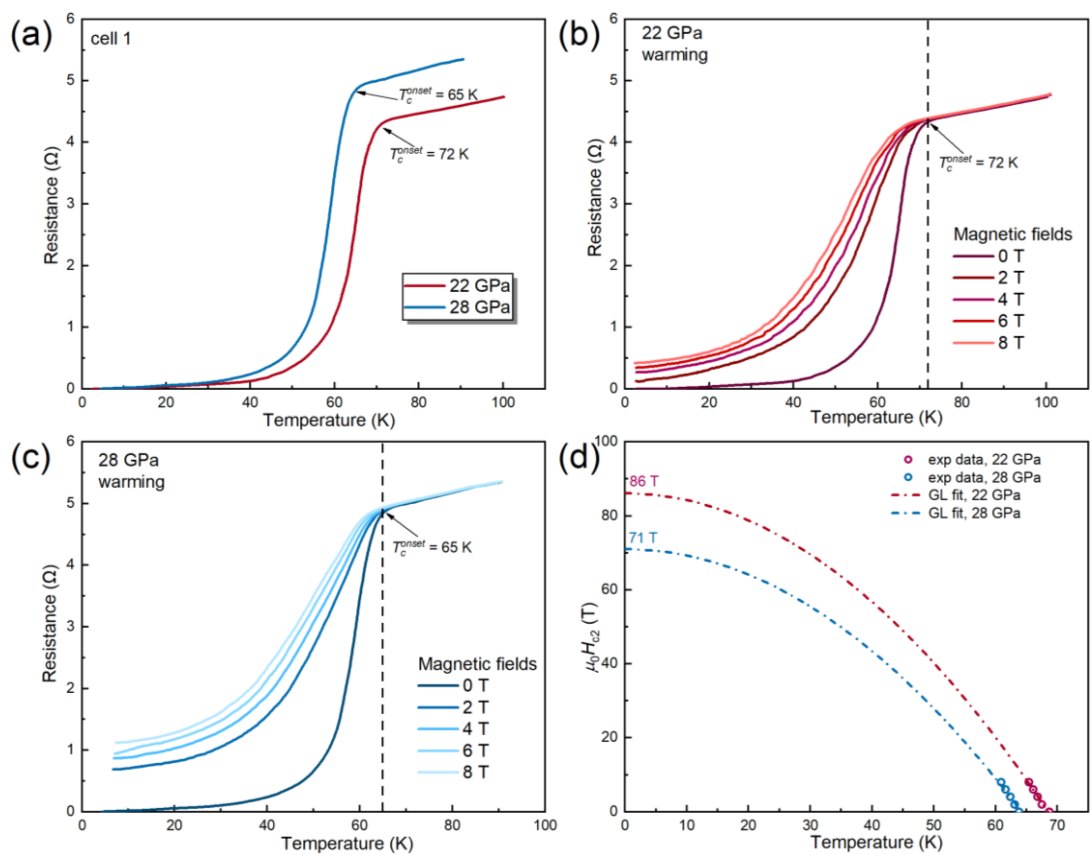

FIG. 2. The transport properties of $La_3Ni_2O_{7-\delta}$ single crystal in different pressures. (a) The temperature dependence of resistance at 22 GPa and 28 GPa, respectively. (b) R(T) curves of $La_3Ni_2O_{7-\delta}$ single crystal under external magnetic fields at 22 GPa. (c) R(T) curves of $La_3Ni_2O_{7-\delta}$ single crystal under external magnetic fields at 28 GPa. (d) The upper critical magnetic fields are fitted by Ginzburg-Landau (GL) equations, and indicated by dashed lines.

As shown in Figures 3(a) and 3(b), a small $La_3Ni_2O_{7-\delta}$ sample was positioned at the center of the sample chamber. Four electrodes were in direct contact with the sample, and a microwave wire was placed near the sample, crossing the diamond culet. Through microscopic images, we identified the sample's location in the confocal scanning image and determined the position of the test points. Additionally, the superconducting regions indicated in Figures 3(b), 4(a) and 5(b) represent localized estimates derived from discrete measurement points rather than a complete survey of all possible superconducting regions in the sample. We performed zero-field cooling (ZFC) measurements at 28 GPa and in the temperature range of 25-100 K, and ODMR spectroscopy was obtained at several points within the sample chamber. The procedure included cooling the sample to approximately 25 K in the absence of an external magnetic field, stabilizing it for a set period, and then applying a fixed external magnetic field ($H_Z$) of ~25 G. The temperature was gradually increased, and ODMR spectra were recorded at various temperatures to capture the magnetic response. Figure 3(c) shows the ODMR spectra of point 2 at 28 GPa revealing a notable increase in splitting at 51 K compared to 23 K.

The ODMR splitting values extracted from the field heating (FH) experiments are displayed in Figure S7. To isolate the magnetic response from the effects of stress [23], ODMR spectra were also recorded under zero-field condition [36]. The corresponding magnetic field ($B_Z$) values, derived from the splitting, as shown in Figure 3(d). At points 2 and 4, $B_Z$ increased with temperature up to $T_c$ (~60 K) before plateauing, indicative of a transition from the superconducting to the normal state. The superconducting temperature is consistent with the electrical results in Figure 2(c). In contrast, the $B_Z$ at points 1 and 3 remain unchanged with temperature, suggesting that the sample is not homogeneous. These findings demonstrate that the magnetic field sensed by NV centers varies spatially, reflecting inhomogeneities in the superconducting state, emphasizing the need for spatially resolved measurements to fully characterize the superconducting nature of $La_3Ni_2O_{7-\delta}$ under high pressure. According to the recent reported studies, both sample inhomogeneity and pressure inhomogeneity play crucial roles in the measurement of superconducting transitions[37,38]. The zero-field splitting data demonstrate comparable stress levels in both superconducting and non-superconducting regions. These results indicate that the spatial variation in superconducting transitions arises mainly from sample inhomogeneity rather than stress distribution.

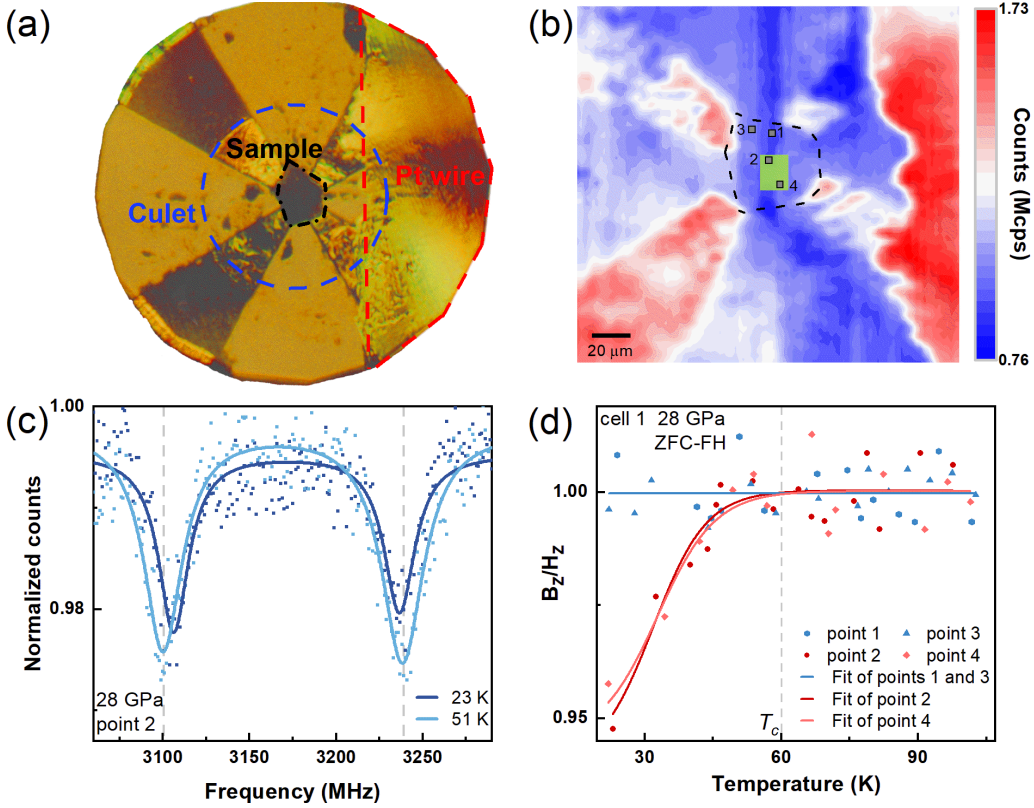

FIG. 3. Local diamagnetism of the $La_3Ni_2O_{7-\delta}$ sample in cell 1 at 28 GPa. (a) Optical microscopic image of the culet surface, including the sample, electrodes and Pt wire. (b) Fluorescence scanning confocal image of the corresponding sample at 28 GPa, four points (1, 2, 3, 4) are selected based on their position with respect to the $La_3Ni_2O_{7-\delta}$ sample. And the green rectangle represents the superconducting region obtained based on the experimental results. (c) ODMR spectra of NV centers at point 2 at two representative temperatures. (d) The ratio of local magnetic field ($B_Z$), extracted from ODMR splitting to a fixed external magnetic field ($H_Z$) of four points.

To increase the $T_c$ of the sample, local diamagnetism measurements were performed at 22 GPa. The change in pressure led to slight variations in both the position and the Z-axis of the sample. Therefore, we selected multiple sample points for measuring under different pressures and presented the points with diamagnetic changes. The detection region, shown in the fluorescence scanning confocal image in Figure 4(a), was along the microwave wire, with measurement points selected along a 40-micron-long gray dashed. A permanent magnet near the cryostat was used to apply the external magnetic field, and point 5, positioned away from the sample, served as a reference to verify the magnetic field strength on the anvil surface. The sample was first cooled to approximately 32 K in zero-field environment, then an external magnetic field of ~20 G was applied. ODMR spectral measurements were conducted during the field-heating and field-cooling processes. Splitting values were analyzed, with stress effects systematically removed during data processing for accuracy.

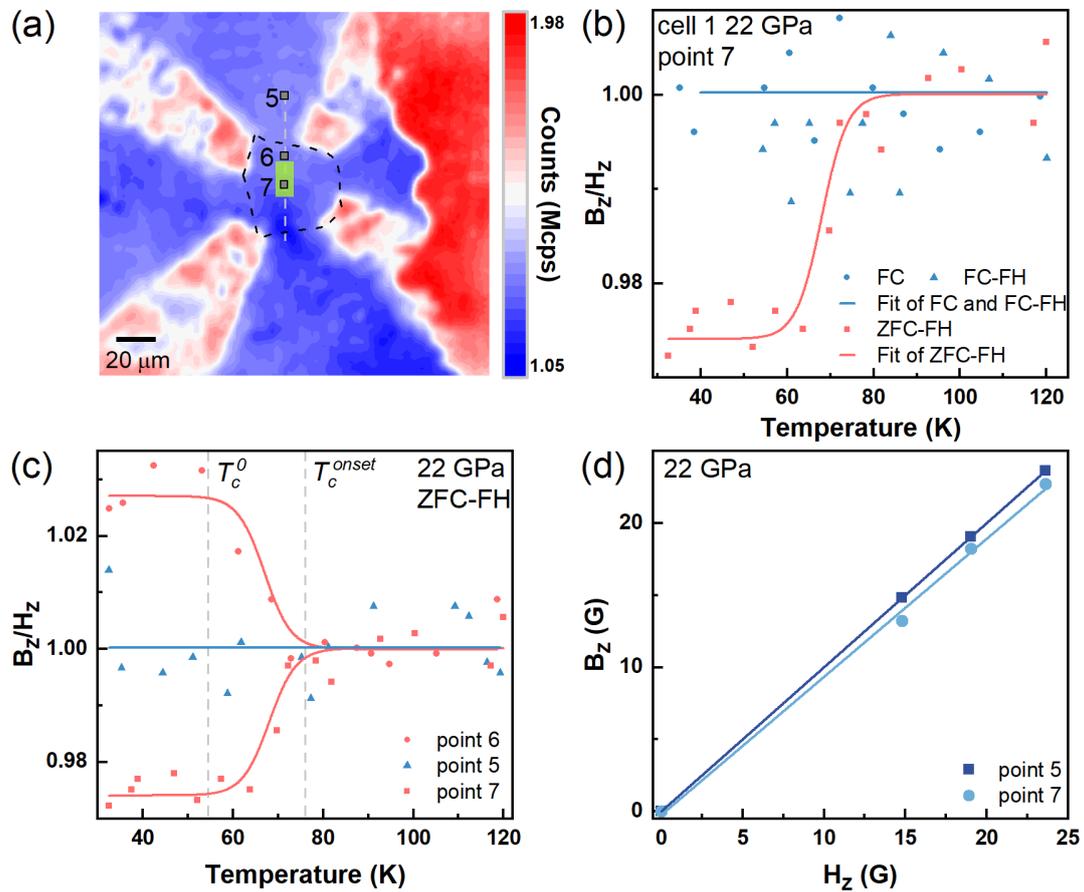

FIG. 4. Local diamagnetism of the $La_3Ni_2O_{7-\delta}$ sample in cell 1 at 22 GPa. (a) Fluorescence scanning confocal image of the sample at 22 GPa, with ODMR spectral measurements conducted at points along the gray dotted line. And the green rectangle represents the superconducting region obtained based on the experimental results. (b) The ratio of $B_Z$ to $H_Z$ at point 7 for three processes: zero-field cooling - field heating

(ZFC-FH), field cooling (FC), and field cooling - field heating (FC-FH). (c) The ratio of $B_Z$ to $H_Z$ for three points collected from the ZFC-FH process. (d) The magnetic field change data for points 5 and 7 extracted from ODMR spectra at 32 K, with point 5 used as a reference for normalization since it is far from the sample.

A series of ODMR spectra were measured at point 7 according to the experimental scheme. The experimental results of ZFC-FH process show that the superconducting transition starts around 76 K and ends at approximately 55 K at 22 GPa. Recently, Wen et al [39]. measured the local diamagnetism of bilayer nickelate at 22 GPa, but did not detect $T_c^0$, likely because they used a polycrystalline sample, whereas we used a single-crystal sample. The electrical data in Figure 2(b) shows that the $T_c$ is about 72 K, which is consistent with the ODMR results. Figure 4(b) shows the results of three experimental processes at point 7, where the $B_Z$ extracted from ODMR splitting [36] remains constant with temperature in both the FC process and the FC-FH process. In cell 2, we increased the external magnetic field, observed the superconducting transition of the FC-FH process, and discussed the reasons [36].

Figure 4(c) displays the ratio of $B_Z$ to $H_Z$ from the ZFC-FH process at three representative points. At point 5, distant from the sample, the ratio remains unchanged with temperature. At point 6, near the superconducting part, the ratio decreases as temperature increases during the superconducting transition, as the superconductor repels the magnetic flux, causing a larger flux at the detection point near the superconductor. Point 7, a clear superconducting region, shows an increase in ratio as the temperature rises. This is apparently due to the repulsion of the magnetic flux by superconductivity. When the sample is cooled below $T_c$, the superconducting region expels magnetic flux, resulting in a reduction of flux directly above the sample and flux accumulation at the edges. The change in splitting at both points is about 3 MHz [36]. Importantly, the opposite behavior at points 6 and 7 confirms that this is not an ordinary magnetic phase transition, but rather a superconducting transition. These findings offer unambiguous proof of superconductivity in nickel-based systems, addressing and resolving the ongoing controversy in the field.

The magnetic field ($B_Z$) sensed by the NV center at points 5 and 7, under different external magnetic fields ($H_Z$), revealed that point 7 had a lower magnetic field than point 5, indicating local diamagnetism at point 7. By comparing the sensed magnetic field with the applied magnetic field, a magnetic field variation diagram was generated, and the slope ($S=\delta B_Z/\delta H_Z$) was calculated for each point as shown in Figure 4(d). The results show a 1 G difference in local diamagnetism between point 7 and point 5 under a 20 G external magnetic field, consistent with the data in Figure 4(c).

To characterize the superconducting transition occurring during the FC process, a new high-pressure sample cell 2 was meticulously prepared. As shown in Figures 5(a) and (b), a series of points were chosen within the region of the second sample. Among these, the FC curve at point 8 was successfully measured, which served as a critical data set for further analysis. Leveraging the obtained test results, the superconducting region was carefully demarcated in Figure 5(b). Specifically, as depicted in Figure 5(b), the superconducting region is delineated by a green rectangular area, providing a clear visual representation of the identified superconducting domain. The detailed experimental outcomes are presented in Figure 5(c). Evidently, the ZFC curve at point 8 exhibits a discernible trend of superconducting transition when subjected to an external magnetic field of approximately 45 G. As the temperature ascends, the magnetic field detected by the NV centers at point 8 gradually increases until reaching the $T_c$. Beyond $T_c$, the magnetic field sensed by the NV centers at this point remains constant.

To conduct a more comprehensive and accurate assessment of the magnetic field evolution during the FC process, the applied magnetic field was deliberately augmented to around 47 G for subsequent testing. This adjustment proved to be fruitful, as it enabled the successful measurement of the magnetic field variations experienced by the NV centers during both the ZFC and FC processes at point 8. These measurements have been instrumental in elucidating the local diamagnetic properties of the single - crystal sample, thereby contributing significantly to the understanding of its superconducting behavior. Moreover, the experimental demonstrations of the ZFC and FC processes at two additional points are displayed in the Supplemental Material Fig. S10 [36].

In addition, by varying the external magnetic field, the ODMR spectra were measured at point 8 within the sample region and at point 9 outside the sample region. Through extracting the magnetic fields sensed at these two points, we obtained the magnetic field variations depicted in Figure 5(d). Using the magnetic field at point 9 as a reference, we observed that the magnetic field change at point 8 is less pronounced than that at point 9. This observation is consistent with the characteristic of point 8 being a superconducting state.

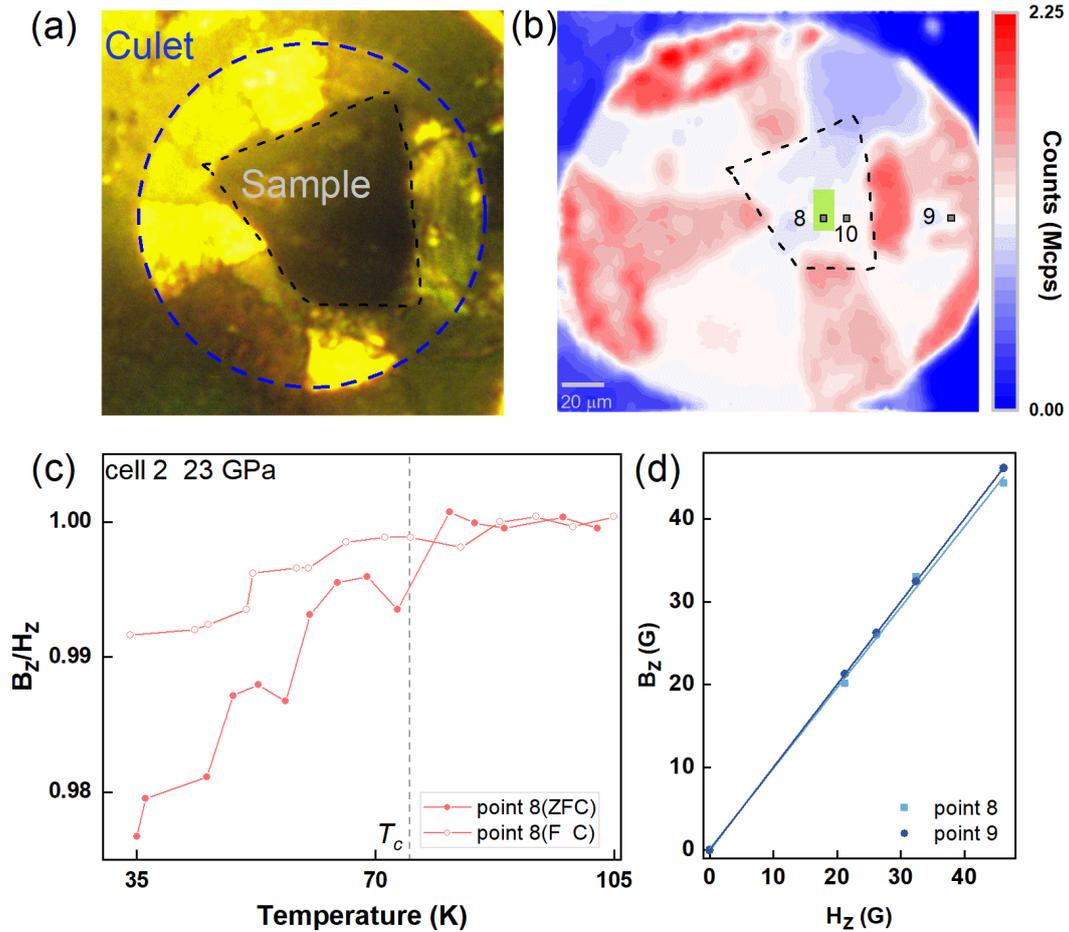

FIG. 5. Local diamagnetism of the $La_3Ni_2O_{7-\delta}$ sample in cell 2 at 23 GPa. (a) Optical microscopic image of the culet surface at 23 GPa, showing the sample and electrodes. The sample chamber is outlined by a blue dashed circle, and the sample is black region (same as in b). (b) Fluorescence scanning confocal image of the sample at 23 GPa, with ODMR spectral measurements performed at selected points (points 8, 9). (c) Ratio change of $B_Z$ to $H_Z$ at four points during the ZFC-FH and FC-FH processes. Solid icons represent ZFC-FH, and hollow icons represent FC-FH for each point. (d) The magnetic field change curves for points 10 and 11 extracted from ODMR spectra at 35 K, with point 11 used as a reference for normalization since it is far from the sample.

The observed diamagnetic response is only 2-5%. We note this value should not be taken as the superconducting volume fraction. We attribute the relative low diamagnetic response to two aspects: Firstly, since the electrical and magnetic data of the sample will be measured simultaneously, as shown in Figure S1, our sample is not in close contact with the diamond surface containing the NV centers. Secondly, the oxygen content has a relatively significant influence on the superconducting properties of nickel-based oxides [40]. For the $La_3Ni_2O_{7-\delta}$ sample, the synthesis process is sensitive to oxygen concentration, making it challenging to produce samples with high superconducting volume fraction. These factors may all lead to relatively weak results. Estimating the superconducting volume fraction accurately remains challenging with current methods. Our NV-center measurements are surface-sensitive, preventing precise quantification of the volume-magnetic field correlation. Spatial sampling limitations further complicate area-based estimates. While Li et al.[37] reported >40% superconducting ratio via susceptibility measurements, this quality-dependent parameter primarily demonstrates Meissner effect evidence rather than intrinsic properties. The method's inherent constraints preclude definitive volume fraction determination.

In this study, we demonstrated zero resistance in the single-crystal $La_3Ni_2O_{7-\delta}$ and examined its local diamagnetism under varying pressures, confirming the presence of superconductivity. Using NV center sensors, we verified the Meissner effect in the sample, indicating its superconducting behavior. Our results show a superconducting transition temperature ($T_c$) of approximately 60 K at 28 GPa, which increases to around 76 K as pressure decreases to 22 GPa. In another high-pressure sample, we successfully measured the ZFC and FC curves of the superconducting transformation of a single crystal sample at 23 GPa. These findings offer unambiguous proof of superconductivity in $La_3Ni_2O_{7-\delta}$ systems, addressing and resolving the ongoing controversy in the field. The use of diamond NV quantum sensing to measure local diamagnetism in inhomogeneous samples demonstrates its effectiveness in challenging environments, such as high pressure and inhomogeneous materials, reaffirming the precision of NV centers in magnetic measurements.

*Acknowledgements*-This work was supported by the National Key R&D Program of China (Grants No. 2022YFA1405500, No. 2023YFA1406500), Innovation Program for Quantum Science and Technology (No. 2024ZD0302100), the Youth Innovation Promotion Association of CAS (No. 2021446), the National Natural Science Foundation of China (No. 12204484, No. 12425404, No. 12474137), Anhui key research and development program(2022h11020007), and the HFIPS Director's Fund of Chinese Academy of Sciences (Nos. BJPY2023B02). We thank the research group led by N. Y. Yao for their support of the NV center implantation technology.

*Note added*. During the revision of our manuscript, we became aware of a recently published study [39] addressing similar aspects of this research topic.